\documentclass{nature}

\usepackage{lettrine}
\usepackage{graphicx}
\usepackage{dcolumn}
\usepackage{bm}
\usepackage{amssymb}
\usepackage{amssymb}
\usepackage{epsfig}
\usepackage[10pt]{moresize}
\usepackage{comment}
\def\>{\rangle}

\title{Optomechanically-induced transparency in parity-time-symmetric microresonators }

\author{H. Jing$^{1,2\star}$, \c{S}ahin K. \"{O}zdemir$^{3\dag}$, Z. Geng$^{1}$, Jing Zhang$^{4}$, Xin-You L\"{u}$^{2,5}$, Bo Peng$^{3}$, Lan Yang$^{3}$ \& Franco Nori$^{2,6\ddag}$}

\begin{document}

\maketitle

\begin{affiliations}
\item Department of Physics, Henan Normal University,
Xinxiang 453007, P. R. China
 \item CEMS, RIKEN, Saitama, 351-0198, Japan
 \item Electrical and Systems
Engineering, Washington University, St. Louis, Missouri 63130,
USA
 \item Department of Automation, Tsinghua
University, Beijing 100084, P. R. China
  \item School of physics, Huazhong University of Science and
Technology, Wuhan 430074, P. R. China
  \item Physics Department, The University of
Michigan, Ann Arbor, MI 48109-1040,
USA
\\$^\star$e-mail:jinghui73@gmail.com
$^\dag$e-mail:ozdemir@ese.wustl.edu
$^\ddag$e-mail:fnori@riken.jp

\end{affiliations}

\begin{abstract}
Optomechanically-induced transparency (OMIT) and the associated slowing of light provide the basis for storing photons in
nanoscale devices. Here we study OMIT in parity-time ($\mathcal{PT}$)-symmetric microresonators with a tunable gain-to-loss
ratio. This system features a reversed, non-amplifying transparency, i.e., an inverted-OMIT. When the
gain-to-loss ratio is varied, the system exhibits a transition from a $\mathcal{PT}$-symmetric phase to a broken-$\mathcal{PT}$-symmetric phase. This $\mathcal{PT}$-phase transition results in the reversal of the pump and gain dependence of the transmission rates. Moreover, we show that by tuning the pump power at a fixed gain-to-loss ratio, or the gain-to-loss ratio at a fixed pump power, one can switch from slow to fast light and vice versa. These findings provide new tools for controlling light propagation using nanofabricated phononic devices.
\end{abstract}

\lettrine[lines=2]{R}ecent advances in steering a macroscopic mechanical object in the deep quantum regime\cite{Cleland,Cleland2,Cleland3} have motivated theoretical studies to understand the physics of photon-phonon interactions in cavity optomechanics (COM), and also led to exciting experimental studies on quantum nanodevices\cite{OM-review,OM-review2}. In particular, the experimental demonstration of OMIT allows the control of light propagation at room temperature using nano- and micro-mechanical structures\cite{OMIT-single,OMIT-single2,OMIT-single3}. The underlying physics of OMIT is formally similar to that of electromagnetically-induced transparency (EIT) in three-level $\Lambda$-type atoms\cite{OMIT-single,OMIT-single2,OMIT-single3,EIT-Harris,EITanalog} and its all-optical analogs demonstrated in various physical systems\cite{EITanalog,Petr}. The resulting slow-light propagation provides the basis for a wide range of applications\cite{EIT-Harris}. Mechanically-mediated delay (slow-light) and advancement (fast-light) of microwave pulses were also demonstrated in a superconducting nanocircuit\cite{EMIT,EMIT2,slow-fast,slow-fast2}. These experimental realizations offer new prospects for on-chip solid-state architectures capable of storing, filtering, or synchronizing optical light propagation.

As a natural extension of single-cavity structures, COM with an auxiliary cavity (compound COM: two cavities) has also attracted intense interest. The interplay between COM interactions and tunable optical tunnelling provides a route for implementing a series of important devices, such as phonon lasers\cite{phonon-laser}, phononic processors for
controlled gate operations between flying (optical) or stationary (phononic) qubits\cite{processor,processor2}, and coherent optical wavelength converters\cite{transducer,transducer2,transducer3}. Enhanced nonlinearities\cite{Ludwig} and highly-efficient photon-phonon energy transfer\cite{Fan,OMIT-double,OMIT-double2} are other advantages of the compound COM. These studies were performed with passive (lossy, without optical gain) resonators.

Very recently, an optical system whose behavior is described by $\mathcal{PT}$-symmetric Hamiltonians (i.e., the commutator $[H,\mathcal{PT}]=0$)\cite{PT1,PT1-2} was demonstrated in a system of two coupled microresonators, one of which has passive loss and the other has optical gain (active resonator)\cite{Sahin1}. Observed features include: real eigenvalues in the $\mathcal{PT}$-symmetric regime despite the non-Hermiticity of the Hamiltonian, spontaneous $\mathcal{PT}$-symmetry breaking, as well as complex eigenvalues and field localization in the broken $\mathcal{PT}$-symmetry regime. Moreover, nonreciprocal light transmission due to enhanced optical nonlinearity in the broken $\mathcal{PT}$-symmetry regime was demonstrated\cite{PT1}. Such a $\mathcal{PT}$-symmetric structure provides unique and previously-unattainable control of light and even sound\cite{PT1,PT1-2,PT2,PT2-2,PT3,Sahin1,Sahin2}. Manipulating the photon-phonon interactions in such systems opens new regimes for phonon lasing and quantum COM control\cite{Jing}.

In this paper, we show that a compound COM with $\mathcal{PT}$-symmetric microresonators leads to previously unobserved features and provides new
capabilities for controlling light transmission in micro- and nano-mechanical systems. Particularly, we show: (i) a gain-induced reversed transparency
(inverted-OMIT), i.e. an optical spectral dip between two strongly-amplifying sidebands, which is in contrast to the non-absorptive peak between strongly
absorptive sidebands in the conventional passive OMIT; (ii) a reversed pump dependence of the optical transmission rate, which is most significant when the
gain and loss are balanced (i.e., optical gain in one subsystem completely compensates the loss in the other); and (iii) a gain-controlled switching from
slow (fast) light to fast (slow) light in the $\mathcal{PT}$-symmetric ($\mathcal{PT}$-breaking) regime, within the OMIT window. These features
of the active OMIT enable new applications which are not possible in passive COM.

The inverted-OMIT observed here in an active COM, composed of a
passive and an active optical microresonator, is reminiscent of
the inverted-EIT observed in all-optical systems, composed of
one active and one passive fiber loop\cite{IEIT}. In Ref.\,\cite{IEIT} a non-amplifying window accompanied with a negative
group delay (fast light) was reported. However, our active OMIT, relies on hybrid photon-phonon interactions in a
compound COM\cite{OMIT-single}. In the $\mathcal{PT}$-symmetric
regime, it provides the first OMIT analog of the optical
inverted-EIT\cite{IEIT}. Distinct features of the inverted-OMIT
that cannot be observed in the optical inverted-EIT are also
revealed in the broken-$\mathcal{PT}$-symmetric regime.

\section*{Results}

\subsection{The active COM system.} We consider a system of two coupled whispering-gallery-mode microtoroid resonators\cite{EITanalog,phonon-laser,Peng1,loss-2014}. One of the resonators is passive and contains a mechanical mode of frequency $\omega_\mathrm{m}$ and an effective mass $m$\cite{phonon-laser}. We refer to this resonator as the optomechanical resonator.
The second resonator is an active resonator which is coupled to the first one through an evanescent field. The coupling strength $J$ between
the resonators can be tuned by changing the distance between them. As in Ref.\,\cite{Sahin1}, the active resonator can be fabricated from $\mathrm{Er}^{3+}$-doped silica and can emit photons in the $1550\,\mathrm{nm}$ band, when driven by a laser in the $980\,\mathrm{nm}$ or $1450\,\mathrm{nm}$ bands. The resonators can exchange energies only in the emission band of $1550\,\mathrm{nm}$, so the gain photons can tunnel through the air gap between the resonators and provide a gain $\kappa$ to compensate the optical loss $\gamma$ in the passive resonator\cite{Sahin1}.

Tuning the gain-to-loss ratio, while keeping $J$ fixed, leads to two remarkably distinct regimes, i.e. broken- and
unbroken-$\mathcal{PT}$-symmetry regimes, that are characterized by
distinct normal mode-splitting and linewidths\cite{Sahin1,Jing}.
Our aim here is to study OMIT in these two distinct regimes,
focusing on the role of $\kappa/\gamma$. To this end, as in the
conventional OMIT\cite{OMIT-single}, both a pump laser of
frequency $\omega_\mathrm{L}$ and a weak probe light of frequency
$\omega_p$ are applied (see Fig.\,1). The
field amplitudes of the pump and probe are given by
$
E_\mathrm{L}=({{2P_\mathrm{L}\gamma}/{\hbar\omega_\mathrm{L}}})^{1/2},$
$\varepsilon_p=({{2P_{\mathrm{in}}\gamma}/{\hbar\omega_p}})^{1/2},
$
where $P_\mathrm{L}$ and $P_{\mathrm{in}}$ are the pump and probe powers.

The Hamiltonian of this three-mode COM system can be written as
\begin{eqnarray}
H&=&H_0+H_{\mathrm{int}}+H_{\mathrm{dr}},\nonumber\\
H_0&=&\hbar\Delta_\mathrm{L} (a_1^\dag a_1+a_2^\dag a_2)+\frac{p^2}{2m}+\frac{1}{2}m\omega_\mathrm{m}^2x^2,\nonumber\\
H_{\mathrm{int}}&=&-\hbar J(a_1^\dag a_2+a_2^\dag a_1)-\hbar
ga_1^\dag
a_1x,\nonumber\\
H_{\mathrm{dr}}&=&i\hbar (E_\mathrm{L}a_1^\dag -E_\mathrm{L}a_1+\varepsilon_pa_1^\dag
e^{-i\xi t}-\varepsilon_pa_1e^{i\xi t}),
\end{eqnarray}
where $a_1$ and $a_2$ denote the annihilation operators of the
bosonic fields in the microresonators with resonance frequency
$\omega_c$ and radius $R$, $g={\omega_c}/{R}$ is the COM coupling
rate, $x=x_0(b+b^\dag)$ is the mechanical position operator,
$x_0=[{\hbar}/(2m\omega_\mathrm{m})]^{1/2}$, and $b$ corresponds to the
annihilation operator for the phonon mode. The pump-resonator,
probe-resonator, and probe-pump frequency detunings are,
respectively, denoted by $$\Delta_\mathrm{L}=\omega_c-\omega_\mathrm{L},~~
\Delta_p=\omega_p-\omega_c,~~\xi=\omega_p-\omega_\mathrm{L}.$$

The Heisenberg equations of motion (EOM) of this compound system are ($\hbar=1$)
\begin{eqnarray}
&&\ddot{x}+\Gamma_\mathrm{m}\dot
x+\omega_\mathrm{m}^2x=\frac{g}{m}a_1^+a_1,\\
&&\dot a_1=(-i\Delta_\mathrm{L} +igx-\gamma)a_1+iJa_2+E_\mathrm{L}+\varepsilon_pe^{-i\xi t},\\
&&\dot a_2=(-i\Delta_\mathrm{L} +\kappa)a_2+iJa_1,\label{eqx1}
\end{eqnarray}
where $\Gamma_\mathrm{m}$ is the mechanical damping rate. The optical or mechanical gain and damping terms are added phenomenologically into the EOM\cite{OMIT-double,Peng1,Jing}; they can also be incorporated into Eq.\,(1), resulting in a non-Hermitian Hamiltonian\cite{PT1,PT1-2,PT2,PT2-2,PT3,Sahin1,Sahin2,Jing,loss-2014}. We note that $\kappa<0$ in Eq.\,(\ref{eqx1}) corresponds to a passive-passive
COM; thus $\kappa/\gamma<0$ and $\kappa/\gamma>0$ define, respectively, a passive-passive COM and a passive-active COM.

The steady-state values of the dynamical variables are
\begin{eqnarray}
x_s&=&\frac{g}{m\omega_\mathrm{m}^2}|a_{1,s}|^2,\\
a_{1,s}&=&\frac{E_\mathrm{L}(i\Delta_\mathrm{L}-\kappa)}{(i\Delta_\mathrm{L}-\kappa)(\gamma+i\Delta_\mathrm{L}-igx_s)+J^2},\\
a_{2,s}&=&\frac{iJE_\mathrm{L}
}{(i\Delta_\mathrm{L}-\kappa)(\gamma+i\Delta_\mathrm{L}-igx_s)+J^2}.
\end{eqnarray}
For $\Delta_\mathrm{L}=0$, by choosing $J^2={\kappa\gamma }$ or
$\kappa/\gamma\rightarrow 1$ for $J=\gamma$, one can identify a gain-induced
transition from the linear to the nonlinear regime that significantly enhances COM interactions\cite{Jing}.
Here we focus on the effects of the gain-loss balance on the OMIT and the associated optical group
delay, which, to our knowledge, has not been studied previously.

We proceed by expanding each operator as the sum of its steady-state value and a small fluctuation around that value, i.e. $a_1=a_{1,s}+
\delta\!a_1,~a_2=a_{2,s}+\delta\!a_2,~x=x_s+\delta\!x$. After eliminating the steady-state values, we obtain the linearized EOM, which can be solved using the ansatz (see the Method)
\begin{eqnarray}
\left(
\begin{array}{c}
  \langle\delta\!a_1\rangle\\
  \langle\delta\!a_2\rangle\\
 \langle\delta\!x\rangle\\
\end{array}
\right) = \left(
\begin{array}{c}
  \delta\!a_{1+}\\
  \delta\!a_{2+}\\
 \delta\!x_+\\
\end{array}
\right) e^{-i\xi t} +
\left(
\begin{array}{c}
  \delta\!a_{1-}\\
  \delta\!a_{2-}\\
 \delta\!x_-\\
\end{array}
\right) e^{i\xi t}.
\end{eqnarray}
The optical fluctuation in the optomechanical resonator $\mathcal{A}\equiv \delta\!a_{1+}$, the quantity of interest here, is
\begin{equation}
\mathcal{A}=\frac{[(\omega_\mathrm{m}^2-\xi^2-i\xi\Gamma_\mathrm{m})
\mathcal{G}_{2}m+ig^2n_1\mu_-]\mu_+\varepsilon_p}{(\omega_\mathrm{m}^2-\xi^2-i\xi\Gamma_\mathrm{m})
\mathcal{G}_{1}\mathcal{G}_{2}m-ig^2n_1(\mathcal{G}_{2}\mu_+
-\mathcal{G}_{1}\mu_-)},~~~
\end{equation}
where $n_1=|a_{1,s}|^2$ is the intracavity photon
number of the passive resonator, $\mu_\pm=-\kappa-i\xi\pm i\Delta_\mathrm{L}$, and
\begin{eqnarray}
\mathcal{G}_{1}&=&(i\Delta_\mathrm{L}+\gamma-igx_s-i\xi)\mu_++J^2, \nonumber\\
\mathcal{G}_{2}&=&(-i\Delta_\mathrm{L}+\gamma+igx_s-i\xi)\mu_-+J^2.
\end{eqnarray}

The expectation value of the output field can then be obtained by using the standard input-output relation, i.e.
$a^{\mathrm{out}}_1(t)=a^{\mathrm{in}}_1-\sqrt{2\gamma}\,a_1(t)$, where $a^{\mathrm{in}}_1$ and $a^{\mathrm{out}}_1(t)$ are the input and output field operators. Then the optical transmission rate $\eta(\omega_p)$ (i.e., the amplitude square of the ratio of the output field amplitude to the input probe field amplitude, $\eta(\omega_p)\equiv\left|t(\omega_p)\right|^2=\left|a^{\mathrm{out}}_1(t)/a^{\mathrm{in}}_1\right|^2$) is
\begin{equation}\label{eqxx}
\eta(\omega_p)=\left|1-(2\gamma/\varepsilon_p)\mathcal{A}\right|^2.
\end{equation}
We computed Eq.\,(\ref{eqxx}) with experimentally-accessible values of the system parameters\cite{Sahin1}  to better understand the behavior of the COM in the presence of gain and loss. These parameters are $R=34.5\, \mu$m, $\omega_c= 1.93\times 10^5\,\mathrm{GHz}$, $\omega_\mathrm{m}=2\pi\times 23.4$ \,MHz, $m=5\times 10^{-11}$\,kg, $\gamma=$6.43\,MHz and $\Gamma_\mathrm{m}= 2.4\times10^5$\,Hz. The quality factors of the optical mode and the mechanical mode in the passive resonator are $Q_c=3\times
10^7$ and $2Q_\mathrm{m}/Q_c=10^{-5}$, respectively. Also $\Delta_\mathrm{L}=\omega_\mathrm{m}$, and thus $\Delta_p\equiv\omega_p-\omega_c=\xi-\omega_\mathrm{m}$. Now we discuss how the gain-loss ratio $\kappa/\gamma$, the coupling strength $J$, and the pump power $P_\mathrm{L}$ affect the OMIT. Note that $\kappa/\gamma$ and $J$ are the tunable system parameters that allow one to operate the system in the broken- or unbroken-$\mathcal{PT}$ regimes.

\subsection{Reversed-gain dependence.} Figure 2 depicts the
effect of $\kappa/\gamma$ on the optical transmission rate. By introducing gain into the second microresonator, one
can tune the system to transit from a conventional OMIT profile,
quantified by a transparency window and two sideband dips, to the
inverted-OMIT profile, quantified by a transmission dip and
two sideband peaks (see Fig.\,2a).

Increasing the loss ratio $\kappa/\gamma<0$ in the
passive-passive COM leads to shallower sidebands. When the amount
of gain provided to the second resonator supersedes its loss and
the resonator becomes an active one (amplifying resonator),
increasing $\kappa/\gamma>0$ helps to increase the heights of the sideband peaks until
$\kappa/\gamma=1$, where $\eta$ at sidebands is maximized. Increasing
the gain further leads to the suppression of both the sideband peaks and the on-resonance ($\Delta_p=0$)
transmission (Fig.\,2b). This is in stark
contrast with the observation of monotonically-increasing sideband
peaks in the all-optical EIT system of Ref.\,\cite{IEIT}.

This can be intuitively explained as follows. Under the condition
of $J/\gamma=1$, the system is in the ${\mathcal PT}$-symmetric
phase for $\kappa/\gamma< 1$, whereas it is in the
broken-${\mathcal PT}$ phase for $\kappa/\gamma> 1$. Thus, for
$\kappa/\gamma< 1$, the provided gain compensates a portion of the
losses, which effectively reduces the loss in the system and hence
increases $\eta$\cite{Sahin1}. Increasing the gain above the phase
transition point $\kappa/\gamma=1$ puts the system in the
broken-$\mathcal{PT}$ phase, with a localized net loss
in the passive resonator\cite{Sahin1} (i.e., the field intensity in the
passive resonator is significantly decreased) which reduces the
strength of the COM interactions and hence the value of $\eta$. The
reduction of the transmission by increasing the
gain provides a signature of the $\mathcal{PT}$-breaking regime,
and it is very similar to a recent experiment with two coupled-resonators where it was
shown that increasing (decreasing) the loss of one of the resonators above (below) a
critical level increases (decreases) the intracavity field intensity of the other, enhancing (suppressing) transmission\cite{loss-2014}. Note that increasing (decreasing)
loss is similar to decreasing (increasing) gain. We conclude here
that only in the $\mathcal{PT}$-symmetric regime ($\kappa/\gamma<
1$, with $J/\gamma=1$), the active OMIT can be viewed as an analog
of the optical inverted-EIT\cite{IEIT}.

Figure 3 provides the fine features in the transmission rate, by numerically changed $\kappa/\gamma>0$ by very small steps.
One interesting observation in Fig.\,3 is that when $\kappa=0$, the second resonator has neither gain nor loss, and there still exists OMIT-like spectrum, i.e. small fluctuations around
$\eta\sim 1$. In this case, the coupled-resonator system is still a passive-OMIT because one resonator is lossy while the other has neither loss nor gain. For $\kappa/\gamma=0.01$, we have a small resonance peak (see Fig.\,3a). When the gain is increased, this peak tends to disappear (see e.g. Fig.\,3b) and then evolves into a dip, e.g. $\eta\sim
0$ for $\kappa/\gamma=0.2$ (see Fig.\,3c). This dip can also be manipulated by increasing the gain further [see Fig.\,3(d-f)]. These results imply that one can tune the system from passive-OMIT to active-OMIT, or vice versa, by varying the gain-to-loss ratio  $\kappa/\gamma$. Such transient behaviors have not been revealed previously.

\subsection{Reversed-pump dependence.} For the passive-passive COM, the
transmission rate and the width of the OMIT window increase with
increasing pump power $P_\mathrm{L}$\cite{OMIT-single,OMIT-double} (see
Fig. 4a). For the passive-active COM, where we
observe the inverted-OMIT, increasing the pump power $P_\mathrm{L}$ leads to a significant
decrease of the sideband amplifications (Fig.\,4b).
Here the pump power dependence of the OMIT profile is shown for $\kappa/\gamma=1.5$ (in the $\mathcal{PT}$-breaking regime). We have also performed
 our calculations for $\kappa/\gamma=1$ and $\kappa/\gamma=0.5$, and similarly found that in these cases the sideband amplifications are also reduced as the pump is increased from $P_\mathrm{L}=10\,\mu\mathrm{W}$ to $20\,\mu\mathrm{W}$ (not shown here). Nevertheless, the sideband amplification always reaches its maximum value at the gain-loss balance (see also Fig.\,2b). We note that the counterintuitive effect of reversed pump dependence was also previously demonstrated in coupled optical systems (i.e., no phonon mode was involved) operating at the exceptional point\cite{EP,loss-2014}.

In addition, we have also studied the effect of the mechanical damping on the profiles of the conventional and inverted OMIT at different values of the gain-to-loss ratio. We confirmed that the profiles of both the conventional and the inverted OMIT are strongly affected (i.e., tend to disappear) by increasing the mechanical damping. This highlights the key role of the mechanical mode in observing OMIT-like phenomena.

\subsection{$\mathcal{PT}$-breaking fast light.} The light transmitted in an EIT
window experiences a dramatic reduction in its group velocity due
to the rapid variation of the refractive index within the EIT
window, and this is true also for the light transmitted in the
OMIT window in a conventional passive optomechanical resonator\cite{OMIT-single}. Specifically, the
optical group delay of the transmitted light is given by
\begin{equation}
\tau_g=\frac{d\arg[t(\omega_p)]}{d\omega_p}\bigg|_{\omega_p=\omega_c}.
\end{equation}
We have confirmed that OMIT in the
passive-passive COM leads only to the slowing (i.e., positive
group delay: $\tau_g>0$) of the transmitted light, and that when
the coupling $J$ between the resonators is weak the reduction in
the group velocity approaches to that experienced in a single
passive resonator\cite{OMIT-single,OMIT-double}. In contrast, in
the active-passive COM, one can tune the system to switch from
slow to fast light, or vice versa, by controlling
$P_{\rm L}$ or $\kappa/\gamma$, such that
the COM experiences the $\mathcal{PT}$-phase transition (see Fig.\,5).

In the regime $\kappa/\gamma <1$, as $P_{\rm L}$ is increased from
zero, the system first enters into the slow-light regime
($\tau_g>0$), and $\tau_g$ increases until its peak value. Then it
decreases, reaching $\tau_g=0$, at a critical value of
$P_{\rm L}$ (Fig.\,5a). The higher is the $\kappa/\gamma$, the sharper is the decrease. Increasing $P_{\rm L}$ beyond this
critical value completes the transition from slow to fast light
and $\tau_g$ becomes negative ($\tau_g<0$). After this transition,
the advancement of the pulse increases with increasing $P_{\rm L}$
until it reaches its maximum value (more negative $\tau_g$).
Beyond this point, a further increase in $P_{\rm L}$, again,
pushes $\tau_g$ closer to zero.

In the regime $\kappa/\gamma>1$, increasing $P_{\rm L}$ from zero
first pushes the system into the fast-light regime and increases
the advancement of the pulse ($\tau_g<0$) until the maximum
advance is reached (Fig.\,5b). After this point, the
advance decreases with increasing $P_{\rm L}$ and finally
$\tau_g$ becomes positive, implying a transition to slow light. If
$P_{\rm L}$ is further increased, $\tau_g$ first increases until
its peak value, and then decreases approaching $\tau_g=0$.

The $P_{\rm L}$ value required to observe the transition from
slow-to-fast light (when $\kappa/\gamma<1$) or from fast-to-slow
light (when $\kappa/\gamma>1$) depends on the gain-to-loss ratio
$\kappa/\gamma$ if the coupling strength $J$ is fixed (Fig. 5). This implies that, when $P_{\rm L}$ is kept fixed,
one can also drive the system from slow-to-fast or fast-to-slow
light regimes by tuning $\kappa/\gamma$. A simple picture can be
given for this numerically-revealed feature: for $\Delta_\mathrm{L}\sim 0$,
$\xi\sim 0$, we simply have $\mathcal{G}_1= \mathcal{G}_2^*\sim
(J^2-\kappa\gamma)+i\kappa gx_s$, which is minimized for
$J^2=\kappa\gamma$, or $\kappa/\gamma=1$, $J/\gamma=1$; therefore,
in the vicinity of the gain-loss balance, the denominator of
$\mathcal{A}$ is a real number, and
$\mathrm{Im}(\mathcal{A})/\mathrm{Re}(\mathcal{A})\sim
(1-\gamma/\kappa)^{-1}$, i.e. having reverse signs for
$\kappa/\gamma>1$ or $\kappa/\gamma<1$. Correspondingly,
$\mathrm{arg}(\mathcal{A})$ or $\mathrm{arg}[t(\omega_p)]$ and
hence its first-order derivative $\tau_g\sim (\gamma/\kappa-1)$ (for $J/\kappa=1$). Clearly, the sign of $\tau_g$ can be reversed by tuning from the $\mathcal{{PT}}$-symmetric regime (with $\kappa/\gamma<1$) to the broken-$\mathcal{PT}$ regime (with $\kappa/\gamma>1$). We note that the appearance of the fast light in the $\mathcal{PT}$-breaking regime, where the gain becomes to exceed the loss, is reminiscent of that observed in a gain-assisted or inverted medium\cite{fast}.

In order to better visualize and understand how the switching from the slow-to-fast light and vice versa takes place, when the gain-to-loss ratio
$\kappa/\gamma$ is tuned at a fixed-pump power $P_\mathrm{L}$, or when $P_\mathrm{L}$ is tuned at a fixed value of $\kappa/\gamma$, we present the phase of the transmission
function $t(\omega_p)$ in Fig.\,6(a-c). For this purpose, we choose the values of $P_\mathrm{L}$ and $\kappa/\gamma$ from Fig.\,5, where their
effects on the optical group velocity $\tau_g$ were presented. These calculations clearly show that, near the resonance point ($\delta_p=0$), the slope of the
curves can be tuned from positive to negative or vice versa, by tuning $P_\mathrm{L}$ or $\kappa/\gamma$, which agrees well with the slow-fast light transitions (see Fig.\,5). In sharp contrast, Fig.\,6d shows that for the passive-passive COM (e.g., $\kappa/\gamma=-1$), no such type of sign reversal can
be observed for the slope of the phase curves, corresponding to the fact that only the slow light can exist in that specific situation.

\section*{Discussion}

In conclusion, we have studied the
optomechanically-induced-transparency (OMIT) in
$\mathcal{PT}$-symmetric coupled microresonators with a tunable gain-loss
ratio. In contrast to the conventional OMIT in passive
resonators (a transparency peak arising in the otherwise
strong absorptive spectral region), the active OMIT in
$\mathcal{PT}$-symmetric resonators features an inverted
spectrum, with a transparency dip between two sideband peaks,
providing a COM analog of the all-optical inverted-EIT\cite{IEIT}.
For this active-OMIT system, the counterintuitive effects of gain- or pump-induced suppression of the optical transmission rate are revealed.
In particular, the transition from slow-to-fast regimes by tuning the
gain-to-loss ratio or the pump power is also demonstrated. The possibility of observing the $\mathcal{PT}$-symmetric fast light, by tuning the gain-to-loss
ratio of the coupled microresonators\cite{Sahin1}, has not studied previously. These exotic features of OMIT in $\mathcal{PT}$-symmetric resonators greatly
widens the range of applications of integrated COM devices for
controlling and engineering optical photons. In addition, our work can be extended to study e.g. the OMIT in a quasi-$\mathcal{PT}$ system\cite{loss-2014}, the OMIT cooling of mechanical motion\cite{OMIT-cooling,OMIT-cooling2}, the active-OMIT with two mechanical modes\cite{transducer}, or the gain-assisted nonlinear OMIT\cite{Clerk,Clerk2,Clerk3}.

\begin{methods}

\subsection{Derivation of the optical transmission rate.}

Taking the expectation of each operator given in Eqs.\,(2)-(4), we find the linearized Heisenberg equations as
\begin{eqnarray}
&&<\delta\!\dot a_1>=-(i\Delta_\mathrm{L}+\gamma-igx_s)<\delta\! a_1>+iJ<\delta\! a_2>
+iga_{1,s}<\delta\! x>+\varepsilon_p\exp({-i\xi t}),\nonumber\\
&&<\delta\! \dot a_2>=-(i\Delta_\mathrm{L}-\kappa)<\delta\! a_2>+iJ<\delta\! a_1>,\nonumber\\
&&<\delta\! \ddot x>+\Gamma_\mathrm{m}<\delta\! \dot x>+\omega_\mathrm{m}^2<\delta\! x>=\frac{g}{m}(a_{1,s}^*<\delta\! a_1>+a_{1,s}<\delta\! a_1^+>).
\end{eqnarray}
which can be transformed into the following form, by applying the
ansatz given in Eq.\,(8),
\begin{eqnarray}
&&(i\Delta_\mathrm{L}+\gamma-igx_s-i\xi)\delta\! a_{1+}=iJ\delta\! a_{2+}+iga_{1,s}\delta\! x_++\varepsilon_p,\nonumber\\
&&(i\Delta_\mathrm{L}+\gamma-igx_s+i\xi)\delta\! a_{1-}=iJ\delta\! a_{2-}+iga_{1,s}\delta\! x_-,\nonumber\\
&&(i\Delta_\mathrm{L}-\kappa-i\xi)\delta\! a_{2+}=iJ\delta\! a_{1+},\nonumber\\
&&(i\Delta_\mathrm{L}-\kappa+i\xi)\delta\! a_{2-}=iJ\delta\! a_{1-},\nonumber\\
&&(\omega_\mathrm{m}^2-\xi^2-i\xi\Gamma_\mathrm{m})\delta\! x_+=\frac{g}{m}(a_{1,s}^*\delta\! a_{1+}+a_{1,s}\delta\! a_{1-}^+),\nonumber\\
&&(\omega_\mathrm{m}^2-\xi^2+i\xi\Gamma_\mathrm{m})\delta\! x_-=\frac{g}{m}(a_{1,s}^*\delta\! a_{1-}+a_{1,s}\delta\! a_{1+}^+).
\end{eqnarray}
Solving these algebraic equations leads to
\begin{equation}
\delta\! x_+=\frac{ga_{1,s}^* \mathcal{G}_2\varepsilon_p\mu_+}{(\omega_\mathrm{m}^2 -\xi^2-i\xi\Gamma_\mathrm{m}) \mathcal{G}_1\mathcal{G}_2m-ig^2n_1(\mathcal{G}_2\mu_+-\mathcal{G}_1\mu_-)},\nonumber
\end{equation}
\begin{equation}
\delta\! x_-=\frac{ga_{1,s} \mathcal{G}_2^*\varepsilon_p\mu_+^*}{(\omega_\mathrm{m}^2 -\xi^2+i\xi\Gamma_\mathrm{m}) \mathcal{G}_1^*\mathcal{G}_2^*m+ig^2n_1(\mathcal{G}_2^*\mu_+^*-\mathcal{G}_1^*\mu_-^*)},\nonumber
\end{equation}
\begin{equation}
\delta\! a_{1+}=\frac{[(\omega_\mathrm{m}^2 -\xi^2-i\xi\Gamma_\mathrm{m}) \mathcal{G}_2m+ig^2n_1\mu_-]\mu_+\varepsilon_p}{(\omega_\mathrm{m}^2 -\xi^2-i\xi\Gamma_\mathrm{m}) \mathcal{G}_1\mathcal{G}_2m-ig^2n_1(\mathcal{G}_2\mu_+-\mathcal{G}_1\mu_-)},\nonumber
\end{equation}
\begin{equation}
\delta\! a_{1-}=\frac{ig^2a_{1,s}^{2}\mu_-^*\mu_+^*\varepsilon_p}{(\omega_\mathrm{m}^2 -\xi^2+i\xi\Gamma_\mathrm{m}) \mathcal{G}_1^*\mathcal{G}_2^*m+ig^2n_1(\mathcal{G}_2^*\mu_+^*-\mathcal{G}_1^*\mu_-^*)},\nonumber
\end{equation}
\begin{equation}
\delta\! a_{2+}=\frac{iJ[(\omega_\mathrm{m}^2 -\xi^2-i\xi\Gamma_\mathrm{m}) \mathcal{G}_2m+ig^2n_1\mu_-]\varepsilon_p}{(\omega_\mathrm{m}^2 -\xi^2-i\xi\Gamma_\mathrm{m}) \mathcal{G}_1\mathcal{G}_2m-ig^2n_1(\mathcal{G}_2\mu_+-\mathcal{G}_1\mu_-)},\nonumber
\end{equation}
\begin{equation}
\delta\! a_{2-}=\frac{-Jg^2a_{1,s}^{2}\mu_+^*\varepsilon_p}{(\omega_\mathrm{m}^2 -\xi^2+i\xi\Gamma_\mathrm{m}) \mathcal{G}_1^*\mathcal{G}_2^*m+ig^2n_1(\mathcal{G}_2^*\mu_+^*-\mathcal{G}_1^*\mu_-^*)}
\end{equation}
where we have used $n_i=|a_{i,s}|^2$ $(i=1,\,2)$  and
\begin{eqnarray}
&&\mu_{\pm}=-\kappa-i\xi \pm i\Delta_\mathrm{L},\nonumber\\
&&\mathcal{G}_1=(i\Delta_\mathrm{L}+\gamma-igx_s-i\xi)\mu_++J^2, \nonumber\\
&&\mathcal{G}_2=(-i\Delta_\mathrm{L}+\gamma+igx_s-i\xi)\mu_-+J^2.
\end{eqnarray}

The expectation value $\left<
a_1^{\mathrm{out}}(t)\right>$ of the output field $a_1^{\mathrm{out}}(t)$ can be calculated using the standard input-output relation $a_1^{\mathrm{out}}(t)=a_1^{\mathrm{in}}-\sqrt{2\gamma}a_1(t)$,  where $a_1^{\mathrm{in}}$ and $a_1^{\mathrm{out}}(t)$ are the input and output field operators, and
\begin{equation}
\left<
a_1^{\mathrm{out}}(t)\right>=\left[E_\mathrm{L}/\sqrt{2\gamma}-\sqrt{2\gamma}(a_{1,s}+\delta\!
a_{1-}e^{i\xi t})\right]e^{-i\omega_\mathrm{L}t}+(\varepsilon_p/\sqrt{2\gamma}-\sqrt{2\gamma}\delta\!
a_{1+})e^{-i(\omega_\mathrm{L}+\xi)t}.
\end{equation}
Hence, the transmission rate of the probe field can be written as
$\eta=|t(\omega_p)|^2$, where $t(\omega_p)$ is the ratio of the output field amplitude to the input field amplitude at the probe frequency
\begin{equation}
t(\omega_p)=\frac{\varepsilon_p-2\gamma\, \delta\!
a_{1+}}{\varepsilon_p}= 1-{2\gamma\mathcal{A}/\varepsilon_p},
\end{equation}
where $\mathcal{A}\equiv \delta\!
a_{1+}$ is given in Eq.\,(9). In order to receive some analytical estimations, we take $\omega_\mathrm{m}/\omega_c\sim 0$, $\Delta_{\mathrm{L},p}\sim 0,$ which leads to
$\mu_\pm \sim -\kappa$, $\mathcal{G}_{1,2}\sim J^2-(\gamma\mp igx_s)\kappa$. For $x_s\sim 0$, we have
\begin{equation}\label{dd1}
\eta\simeq \left| 1+\frac{2\kappa\gamma[m\omega_\mathrm{m}^2(J^2-\kappa\gamma)-ig^2n_1\kappa]}{m\omega_\mathrm{m}^2(J^2-\kappa\gamma)^2}  \right|^2,
\end{equation}
i.e. $\eta\sim (J^2-\kappa\gamma)^{-2}$ or $\eta\sim (1-\kappa/\gamma)^{-2}$ (for a fixed value of $J/\gamma=1$). This indicates that the transmission rate $\eta$ tends to be maximized as the gain-to-loss ratio approaches one, that is
$\kappa/\gamma=1$, which was confirmed by our numerical calculations (see Fig.\,2b).

\end{methods}


\begin{thebibliography}{1}

\bibitem{Cleland} O'Connell, A. D. {\it et al.} Quantum ground state and single-phonon control of a mechanical resonator. Nature {\bf 464}, 697 (2010).
\bibitem{Cleland2} Teufel, J. D. {\it et al.} Sideband cooling of micromechanical motion to the quantum ground state. Nature {\bf 475}, 359 (2011).
\bibitem{Cleland3} Chan, J. {\it et al.}  Laser cooling of a nanomechanical oscillator into its quantum ground state. Nature {\bf
478}, 89 (2011).

\bibitem{OM-review} Aspelmeyer, M., Meystre, P. and Schwab, K.  Quantum optomechanics. Physics Today {\bf 65}, 29 (2012).
\bibitem{OM-review2} Aspelmeyer, M., Kippenberg, T. J. and Marquardt, F. Cavity optomechanics. Rev. Mod. Phys. {\bf 86}, 1391 (2014).

\bibitem{OMIT-single}
Safavi-Naeini, A. H. {\it et al.} Electromagnetically induced transparency and slow light with optomechanics. Nature {\bf 472}, 69 (2011).
\bibitem{OMIT-single2}
Weis, S. {\it et al.} Optomechanically induced transparency. Science {\bf 330}, 1520 (2010).
\bibitem{OMIT-single3} Agarwal, G. S. and Huang, S. Electromagnetically induced transparency in mechanical effects of light. Phys. Rev. A {\bf 81}, 041803(R) (2010).

\bibitem{EIT-Harris} Scully, M. O. and Zubairy, M. S. {\it Quantum Optics} (Cambridge University Press, Cambridge, England, 1997).

\bibitem{EITanalog} Peng, B., \"Ozdemir, S. K., Chen, W., Nori, F. and Yang, L. What is and what is not electromagnetically induced transparency in whispering-gallery microcavities. Nature Comm. {\bf 5}, 5082 (2014).

\bibitem{Petr} Anisimov, P. M., Dowling, J. and Sanders, B. C. Objectively discerning Autler-Townes splitting from electromagnetically induced transparency. Phys. Rev. Lett. {\bf 107}, 163604 (2011).
    
\bibitem{EMIT} Xiang, Z.-L., Ashhab, S., You, J. Q. and Nori, F. Rev. Mod. Phys. {\bf 85}, 623 (2013).

\bibitem{EMIT2} Zhou, X. {\it et al.} Slowing, advancing and switching of microwave signals using circuit nanoelectromechanics. Nature Phys. {\bf 9},
179 (2013).

\bibitem{slow-fast} Jiang, C., Chen, B. and Zhu, K. D. Tunable pulse delay and advancement in a coupled nanomechanical resonator-superconducting microwave cavity system. EPL {\bf 94}, 38002 (2011).
\bibitem{slow-fast2} Dong, C., Zhang, J., Fiore, V. and Wang, H. Optomechanically induced transparency and self-induced oscillations with Bogoliubov mechanical modes. Optica {\bf 1}, 425 (2014).

\bibitem{phonon-laser} Grudinin, I. S., Lee, H., Painter, O. and Vahala, K. J. Phonon laser action in a tunable two-level system. Phys. Rev. Lett. {\bf 104}, 083901 (2010).

\bibitem{processor} Stannigel, K. {\it et al.} Optomechanical quantum information processing with photons and phonons. Phys. Rev. Lett. {\bf 109}, 013603 (2012).
\bibitem{processor2} Komar, P. {\it et al.} Single-photon nonlinearities in two-mode optomechanics. Phys. Rev. A {\bf 87}, 013839 (2013).

\bibitem{transducer} Dong, C., Fiore, V., Kuzyk, M. C. and Wang, H. Optomechanical dark mode. Science {\bf 338}, 1609 (2012).
\bibitem{transducer2} Hill, J. T., Safavi-Naeini, A. H., Chan, J. and Painter, O. Coherent optical wavelength conversion via cavity optomechanics. Nature Commun. {\bf 3}, 1196 (2012).
\bibitem{transducer3} Fiore, V., Yang, Y., Kuzyk, M. C., Barbour, R., Tian, L. and Wang, H. Storing optical information as a mechanical excitation in a silica optomechanical resonator. Phys. Rev. Lett. {\bf 107}, 133601 (2011).

\bibitem{Ludwig} Ludwig, M., Safavi-Naeini, A., Painter, O. and Marquardt, F. Enhanced quantum nonlinearities in a two-mode optomechanical system. Phys. Rev. Lett. {\bf 109}, 063601 (2012).

\bibitem{Fan} Fan J. and Zhu, L. Enhanced optomechanical interaction in coupled microresonators. Opt. Express {\bf 20}, 20790 (2012).

\bibitem{OMIT-double}
Yan, X. B. {\it et al.} Coherent perfect absorption, transmission, and synthesis in a double-cavity optomechanical system. Opt. Express {\bf 22}, 4886 (2014).
\bibitem{OMIT-double2}
Jiang, C., Liu, H., Cui, Y. and Li, X. Electromagnetically induced transparency and slow light in two-mode optomechanics. Opt. Express {\bf 21},
12165 (2013).

\bibitem{PT1} Bender C. M. and Boettcher, S. Real spectra in non-Hermitian Hamiltonians having $\mathcal{PT}$ symmetry. Phys. Rev. Lett. {\bf 80}, 5243 (1998).
\bibitem{PT1-2}  Bender, C. M., Gianfreda, M., \"Ozdemir, S. K., Peng, B. and Yang, L. Twofold transition in $\mathcal{PT}$-symmetric coupled oscillators. Phys. Rev. A {\bf 88}, 062111 (2013).

\bibitem{Sahin1} Peng, B. {\it et al.} Parity-time-symmetric whispering-gallery microcavities. Nature Phys. {\bf 10}, 394 (2014).

\bibitem{PT2} Regensburger, A. {\it et al.} Parity-time synthetic photonic lattices. Nature {\bf 488}, 167 (2012).
\bibitem{PT2-2} R\"uter, C. E. {\it et al.} Observation of parity-time symmetry in optics. Nature Phys. {\bf 6}, 192 (2010). 

\bibitem{PT3} Agarwal, G. S. and Qu, K. Spontaneous generation of photons in transmission of quantum fields in $\mathcal{PT}$-symmetric optical systems. Phys. Rev. A {\bf 85}, 031802(R) (2012).

\bibitem{Sahin2} Monifi, F., \"Ozdemir, S. K. and Yang, L. Tunable add-drop filter using an active whispering gallery mode microcavity. Appl. Phys. {\bf 103}, 181103 (2013).


\bibitem{Jing} Jing, H., \"Ozdemir, S. K., L\"u, X.-Y., Zhang, J., Yang, L. and Nori, F.  $\mathcal{PT}$-symmetric phonon laser. Phys. Rev. Lett. {\bf 113}, 053604 (2014).

\bibitem{IEIT}
Oishi, T. and Tomita, M. Inverted coupled-resonator-induced transparency. Phys. Rev. A {\bf 88}, 013813 (2013).

\bibitem{Peng1} Peng, B., \"Ozdemir, S. K., Zhu, J. and Yang, L. Photonic molecules formed by coupled hybrid resonators. Opt. Lett. {\bf 37}, 3435 (2012).

\bibitem{loss-2014} Peng, B., {\it et al.} Loss-induced suppression and revival of lasing. Science {\bf 346}, 328 (2014).

\bibitem{EP} Brandstetter, M. {\it et al.} Reversing the pump-dependence of a laser at an exceptional point. Nature Comm. {\bf 5}, 4034 (2014).

\bibitem{fast} Wang, L. J., Kuzmich, A. and Dogariu, A. Gain-assisted superluminal light
propagation. Nature {\bf 406}, 277 (2000).

\bibitem{OMIT-cooling} Ojanen, T. and Borkje, K. Ground-state cooling of mechanical motion in the unresolved sideband regime by use of optomechanically induced transparency. Phys. Rev. A {\bf 90}, 013824 (2014).
\bibitem{OMIT-cooling2}Guo, Y., Li, K., Nie, W. and Li, Y. Electromagnetically-induced-transparency-like ground-state cooling in a double-cavity optomechanical system. Phys. Rev. A {\bf 90},053841 (2014).

\bibitem{Clerk} Kronwald, A. and Marquardt, F. Optomechanically-induced transparency in the nonlinear quantum regime. Phys. Rev. Lett. {\bf 111}, 133601 (2013).
\bibitem{Clerk2} Lemonde, M.-A., Didier, N. and Clerk, A. A. Nonlinear interaction effects in a strongly driven optomechanical cavity. Phys. Rev. Lett. {\bf 111}, 133602 (2013).
\bibitem{Clerk3} Borkje, K., Nunnenkamp, A., Teufel, J. D. and Girvin, S. M. Signatures of nonlinear cavity optomechanics in the weak coupling regime. Phys. Rev. Lett. {\bf 111}, 133603 (2013).


\end{thebibliography}


\begin{addendum}
 \item[Acknowledgement] We thank Y. Zhang, P. M. Anisimov, and A. Miranowicz for helpful and stimulating discussions. HJ is supported by the NSFC (11274098, 11474087). LY is
supported by ARO Grant No. W911NF-12-1-0026. JZ is supported by the NSFC (61174084, 61134008) and
the NBRPC (2014CB921401). XYL is supported by the NSFC (11374116).
FN is supported by the RIKEN iTHES Project, MURI Center for Dynamic Magneto-Optics, and a Grant-in-Aid for Scientific
Research\,(S).
 \item[Correspondence] Correspondence and requests for materials should be addressed to H.J.
 \item[Author Contributions] H.J. and S.K.O. conceived the idea. H.J. and S.K.O. carried out the research with input from Z.G., J.Z. and X.Y.L. All of the authors discussed the results. H.J., S.K.O. and F.N. wrote the manuscript with comments and refinements from J.Z., X.Y.L. and L.Y.
 \item[Competing Interests] The authors declare that they have no
competing financial interests.
\end{addendum}


\clearpage

\textbf{Figure 1: OMIT in active-passive-coupled
micro-resonators, with a tunable gain-loss ratio.}
The similarity of the passive OMIT and the three-level EIT is well-known\cite{OMIT-single};
in parallel, the active OMIT provides a COM analog of the optical inverted-EIT\cite{IEIT} (see the energy levels with an input gain). Here  $|0_i\rangle=|n_1,n_2,n_\mathrm{m}\rangle$,
$|1_1\rangle=|n_1+1,n_2,n_\mathrm{m}\rangle$,
$|1_2\rangle=|n_1,n_2+1,n_\mathrm{m}\rangle$,
$|1_\mathrm{m}\rangle=|n_1,n_2,n_\mathrm{m}+1\rangle$, while $n_{1,2}$ and $n_\mathrm{m}$
denote the number of photons and phonons, respectively.
\bigskip

\textbf{Figure 2: The transmission rate $\eta$ of the probe light versus the optical detuning
 $\Delta_p=\omega_p-\omega_c$, for different values of the gain-loss ratio.} 
 (a) Around the transient point $\kappa/\gamma=0$. (b) Around the balanced point
 $\kappa/\gamma=1$. The optical  tunnelling rate and the pump power are fixed as $J/\gamma=1$ and $P_\mathrm{L}=10\,\mu\mathrm{W}$, respectively. Note that although it is not seen clearly in (a), the case $\kappa/\gamma=0$ (red dotted line) has OMIT features of sideband dips and a resonance peak (see Fig.\,3).
\bigskip

\textbf{Figure 3: The transmission rate $\eta$ of the probe light in the active-passive system.}
The relevant parameters are taken as $J/\gamma$=1 and $P_\mathrm{L} = 10\,\mu\mathrm{W}.$
\bigskip

\textbf{Figure 4: Diagrams of the transmission rate of the probe light for different systems.} (a) The passive-passive COM system, with $\kappa/\gamma=-1$.
(b) The active-passive COM system, with $\kappa/\gamma=1.5$ (similar results with e.g. $\kappa/\gamma=0.5$ or $\kappa/\gamma=1$ are not shown here).
\bigskip

\textbf{Figure 5: The delay or advance of the probe
 light in active-passive-coupled resonators, characterized by
 positive or negative group delays.} (a) The $\mathcal{PT}$-symmetric regime with $\kappa/\gamma <1$. (b) The
 $\mathcal{PT}$-breaking regime with $\kappa/\gamma\geq 1$. In these calculations we have taken $\Delta_p=0$, $J/\gamma=1$.
\bigskip

\textbf{Figure 6: The phase of the transmission amplitude $t(\omega_p)$ with different parameters.}
(a-b) With different values of pump power $P_\mathrm{L}$. (c) With different values of gain-to-loss ratio $\kappa/\gamma$. The specific values of $P_\mathrm{L}$ and $\kappa/\gamma$ are taken from Fig.\,5, corresponding to the slow and fast light
 regimes. For comparison, the results for the passive-passive COM system are also plotted in (d) with $\kappa/\gamma=-1$.
\bigskip

\clearpage
\begin{figure}
\begin{center}
\epsfig{file=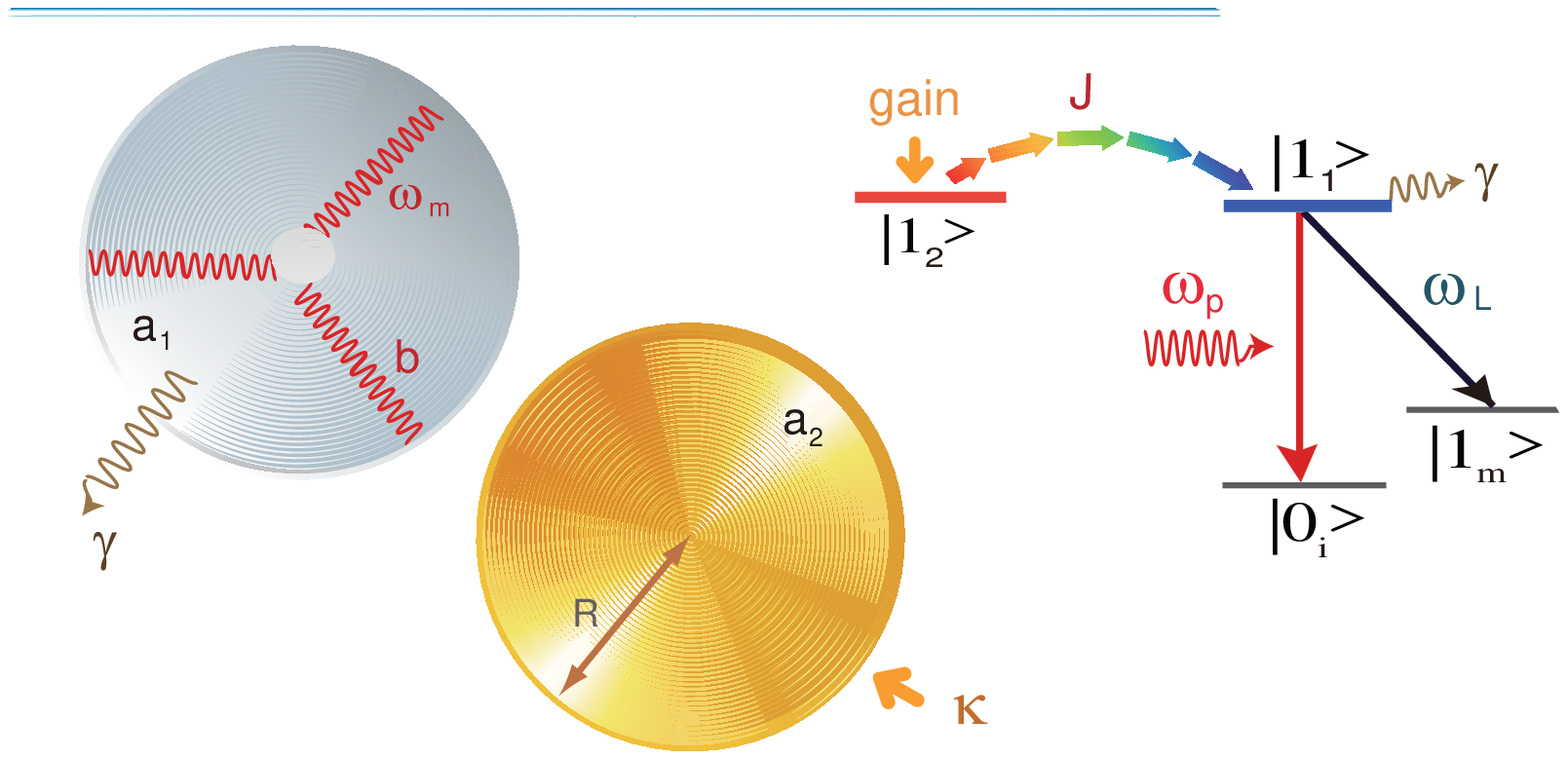,width=12cm}
\end{center}\caption{}
\label{fig1}
\end{figure}

\clearpage
\begin{figure}
\begin{center}
\epsfig{file=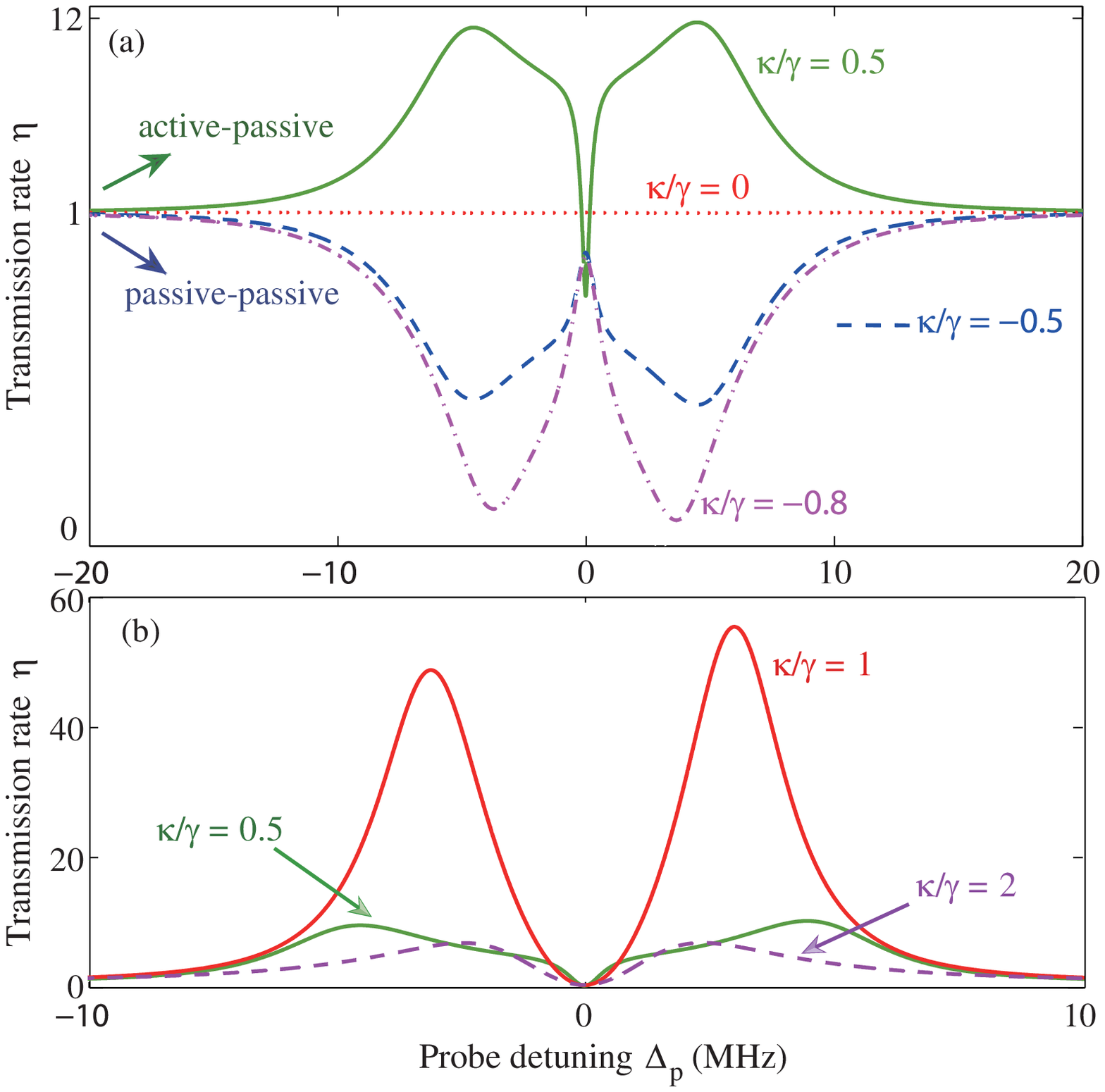,width=12cm}
\end{center}\caption{}
\label{fig2}
\end{figure}

\clearpage
\begin{figure}
\begin{center}
\epsfig{file=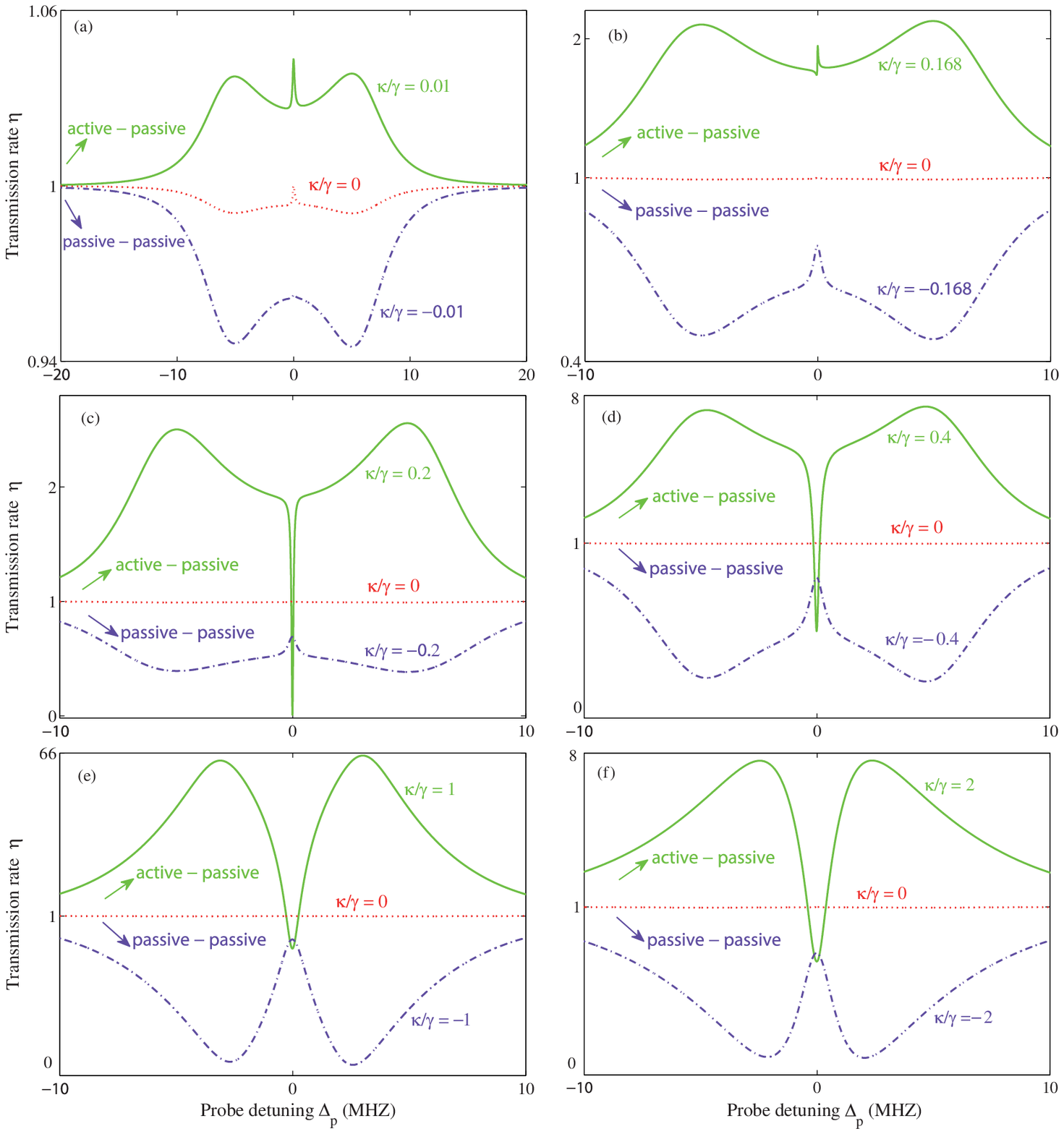,width=15cm}
\end{center}\caption{}
\label{fig3}
\end{figure}
\clearpage

\begin{figure}
\begin{center}
\epsfig{file=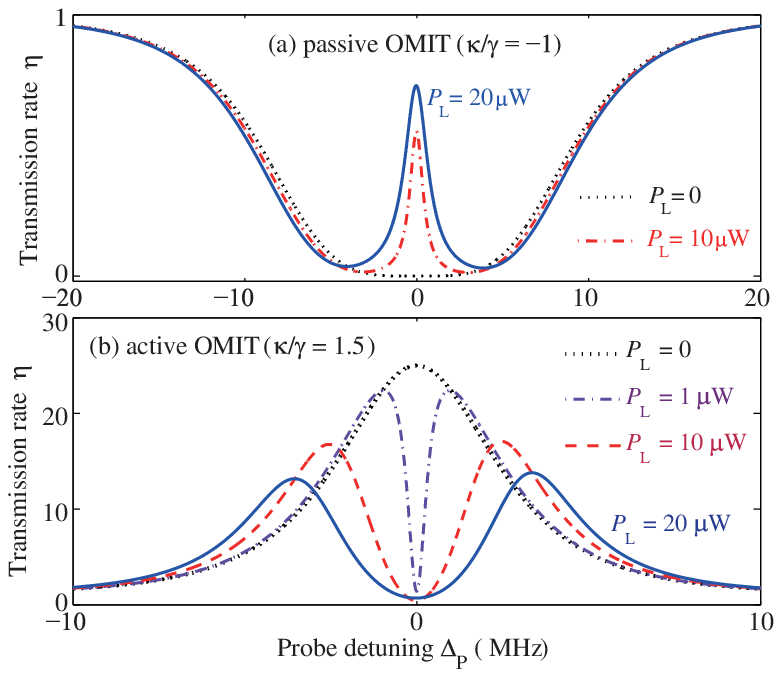,width=12cm}
\end{center}\caption{}
\label{fig4}
\end{figure}
\clearpage

\begin{figure}
\begin{center}
\epsfig{file=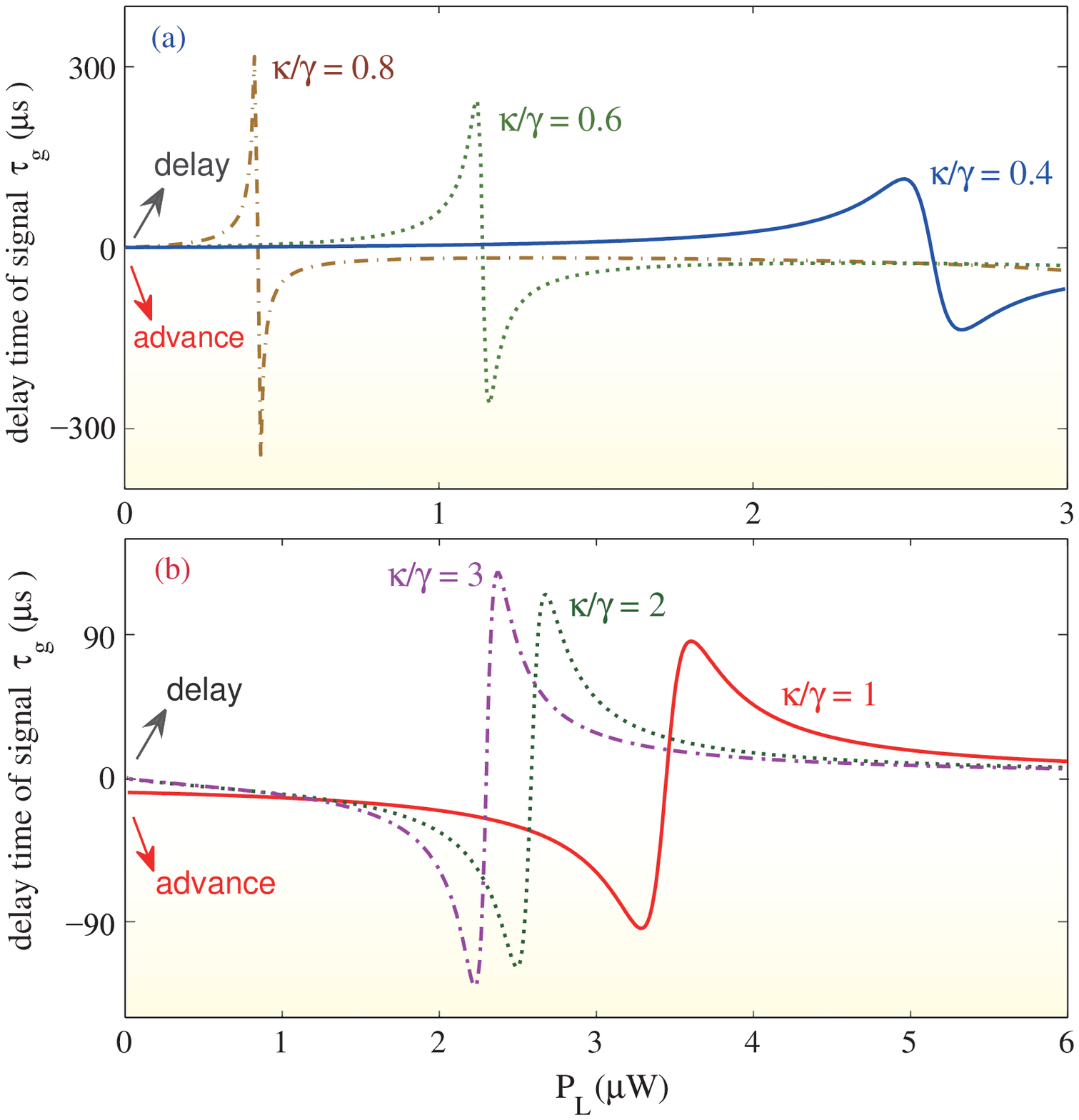,width=12cm}
\end{center}\caption{}
\label{fig5}
\end{figure}
\clearpage

\begin{figure}
\begin{center}
\epsfig{file=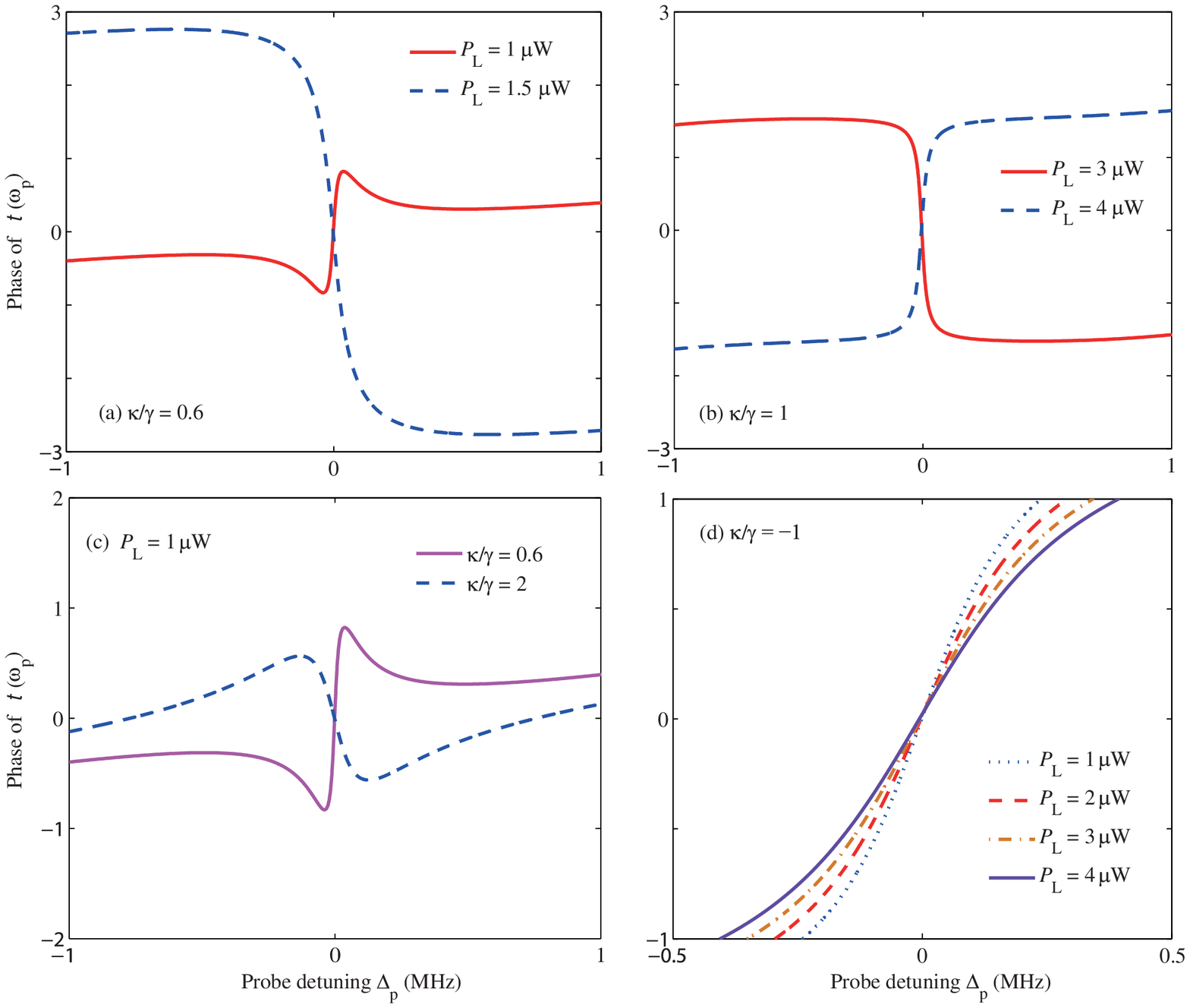,width=15cm}
\end{center}\caption{}
\label{fig6}
\end{figure}

\end{document}